\documentclass[aps,prx,twocolumn,noshowpacs,amsfonts,amssymb,amsmath,longbibliography,superscriptaddress,notitlepage]{revtex4-1}

\usepackage{bm}
\usepackage{graphicx}
\usepackage[caption=false]{subfig}
\begin{document}
%
%
%
%
\title{Smeared phase transitions in percolation on real complex networks}

\author{Laurent \surname{H\'ebert-Dufresne}}
\affiliation{Department of Computer Science and Vermont Complex Systems Center, University of Vermont, Burlington VT}
\affiliation{D\'epartement  de  Physique,  de  G\'enie  Physique,  et  d'Optique, Universit\'e  Laval,  Qu\'ebec  (Qu\'ebec),  Canada  G1V  0A6}
\author{Antoine \surname{Allard}}
\affiliation{D\'epartement  de  Physique,  de  G\'enie  Physique,  et  d'Optique, Universit\'e  Laval,  Qu\'ebec  (Qu\'ebec),  Canada  G1V  0A6}
\date{\today}
\begin{abstract}
Percolation on complex networks is used both as a model for dynamics \textit{on} networks, such as network robustness or epidemic spreading, and as a benchmark for our models \textit{of} networks, where our ability to predict percolation measures our ability to describe the networks themselves. In many applications, correctly identifying the phase transition of percolation on real-world networks is of critical importance. Unfortunately, this phase transition is obfuscated by the finite size of real systems, making it hard to distinguish finite size effects from the inaccuracy of a given approach that fails to capture important structural features. Here, we borrow the perspective of smeared phase transitions and argue that many observed discrepancies are due to the complex structure of real networks rather than to finite size effects only. In fact, several real networks often used as benchmarks feature a smeared phase transition where inhomogeneities in the topological distribution of the order parameter do not vanish in the thermodynamic limit. We find that these smeared transitions are sometimes better described as sequential phase transitions within correlated subsystems. Our results shed light not only on the nature of the percolation transition in complex systems, but also provide two important insights on the numerical and analytical tools we use to study them. First, we propose a measure of local susceptibility to better detect both clean and smeared phase transitions by looking at the topological variability of the order parameter. Second, we highlight a shortcoming in state-of-the-art analytical approaches such as message passing, which can detect smeared transitions but not characterize their nature.
\end{abstract}
\maketitle
%
%
%
%
%
\section{Introduction}
%
Percolation on networks is simple to define \cite{Newman2010}. Given an original network structure, predict the size distribution of connected components if edges (bond percolation) or nodes (site percolation) are randomly removed such that, on average, only a fraction $p$ remain and are said to be ``occupied''. Connected components are groups of nodes that are reachable from one another by following edges, and the relative size of the largest component is a natural order parameter for the connectivity of the system. And while percolation is simple to define, it is widely used to model complex systems. For example, we can model epidemic spreading with percolation by assuming that remaining edges are contacts that would transmit a disease should one of two nodes at its ends become infected \cite{Newman2002}. Because of the breadth and depth of its applications, percolation has become a canonical problem of network science where it reflects the current state of the field: Percolation can be solved on ordered lattices or random networks, but it is much more complicated on the real, complex networks that exist between order and randomness.

One of the most salient features of percolation is its phase transition in infinite systems, i.e., in the thermodynamic limit where the number of nodes formally goes to infinity. When $p$ is close to zero, and few edges or nodes are occupied, the system is almost fully disconnected. As $p$ increases, connected components grow. Eventually, at $p = p_c$, we see the emergence of a macroscopic connected component called the giant connected component (GCC) whose size scales linearly with the size of the system. That is, if we double the size of the system by doubling the number of nodes, there exists a component that will also double in size only if $p>p_c$, while the size of all other components are virtually independent of system size. Therefore, by using the relative size of the GCC as the order parameter, the percolation threshold $p_c$ marks the transition between two phases: (1) A disconnected phase where connectivity does not scale with system size such that the size of the largest connected component vanishes compared to the size of the system; and (2) a connected phase where connectivity scales with system size such that the GCC contains a non-vanishing fraction of all nodes even in an infinite system. In applications, the percolation threshold can, for example, help determine whether a disease can invade a contact network or not.

Predicting the percolation threshold of a complex network is a highly non-trivial task because it depends on the topology of the network at all scales. In fact, even in direct simulations, \textit{detecting} the phase transition can be complicated. While in theory, there is a clean phase transition where our order parameter goes from zero to a non-zero value around $p_c$, in practice this transition is masked by noise caused, for example, by the finite size of real networks. The definition of the GCC involves taking a thermodynamic limit and following its size as we increase the system size, but we typically only have one copy of real systems. For instance, we only have one Internet and it is of fixed sized at any given time. Even if it is growing, the Internet of today is simply not a larger version of the Internet of the nineties but is rather a system with a different structure \cite{leskovec2007graph}.

\begin{figure}
\centering
\includegraphics[width=\linewidth]{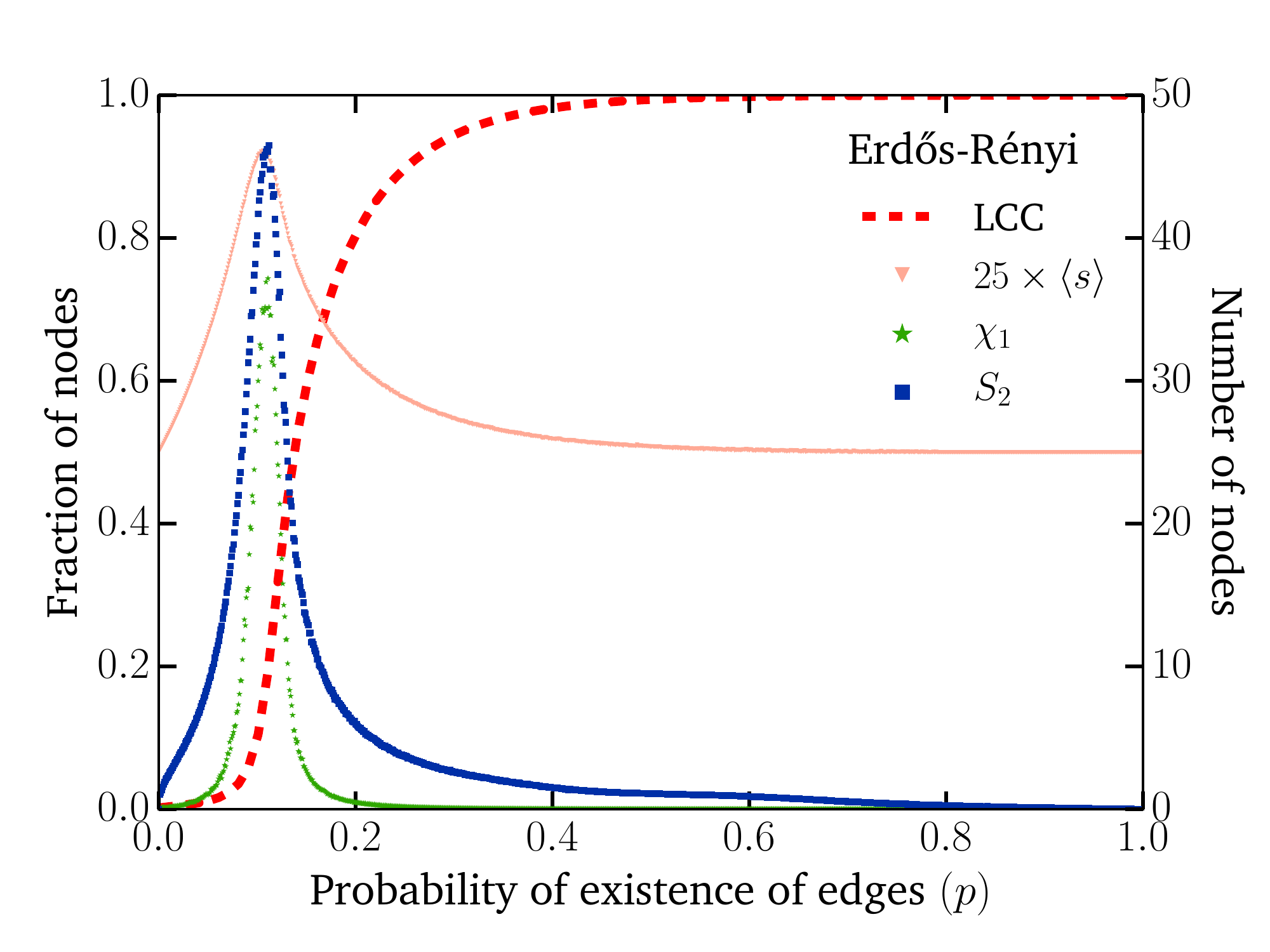}
\caption{\textbf{Detection of the phase transition on a small random network using three common measures.} We study the phase transition, as we vary the probability of existence of edges $p$, in the relative size of the giant connected component (GCC) as approximated by the largest connected component (LCC) when greater than 1\% of network size. We also follow three different metrics --- the susceptibility $\chi_1$ of the LCC, the average size of components smaller than the largest $\langle s \rangle$, and the size of the second largest connected component $S_2$ --- which should all peak at the phase transition. We use a single random realization of an Erd\H{o}s-R\'enyi (ER) random graph of size 1000 with average degree 10. Using a single small network, we capture the typical finite size effects of network structure on the sharpness of the phase transition, despite which the three detection methods all peak together to detect the actual theoretical threshold $p_c = 1/10$.}
\label{fig:theo_detection}
\end{figure}

\begin{figure*}
\centering
\includegraphics[trim={0 1cm 1cm 0},clip,width=0.49\linewidth]{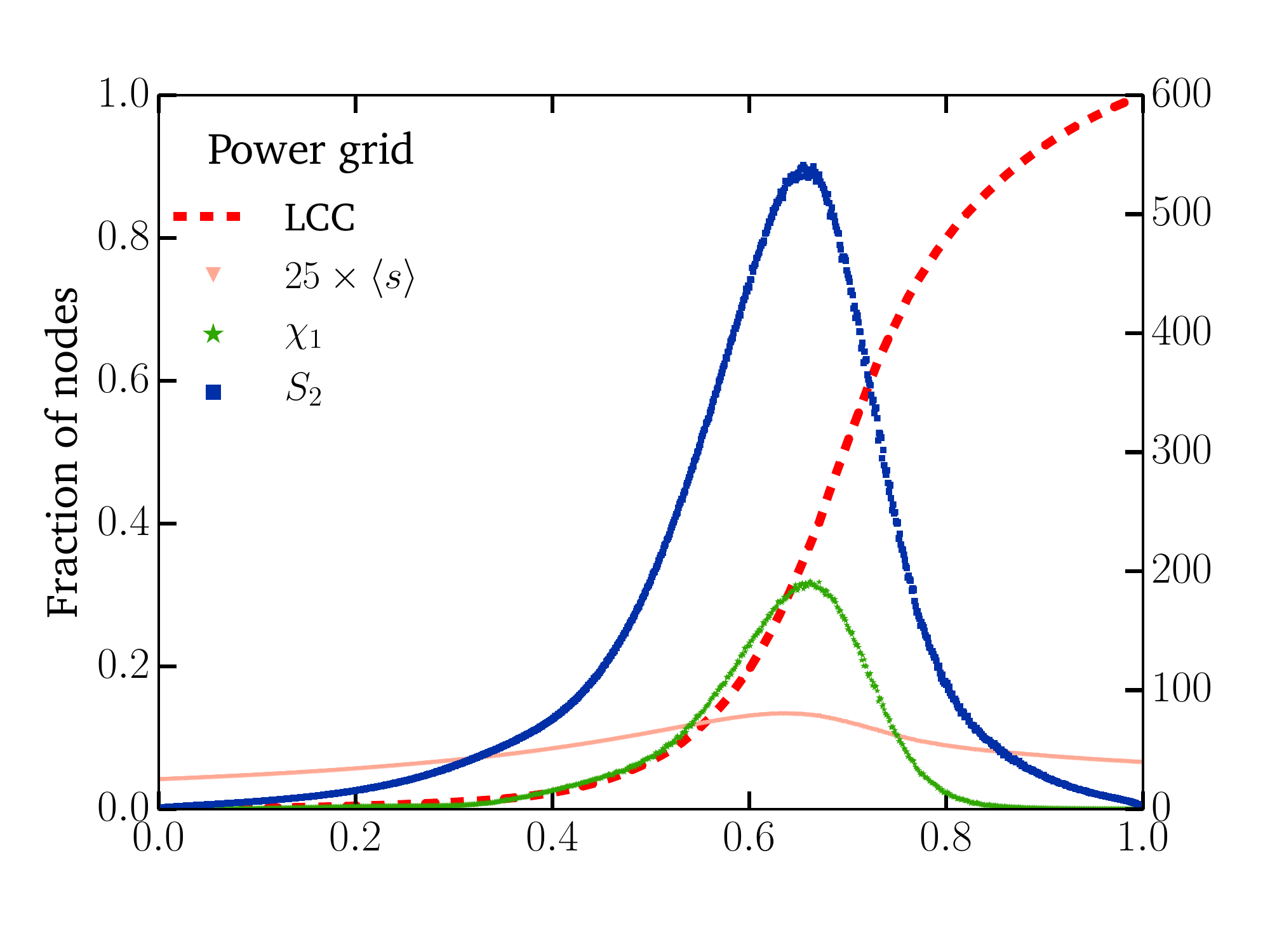}
\includegraphics[trim={1cm 1cm 0 0},clip,width=0.49\linewidth]{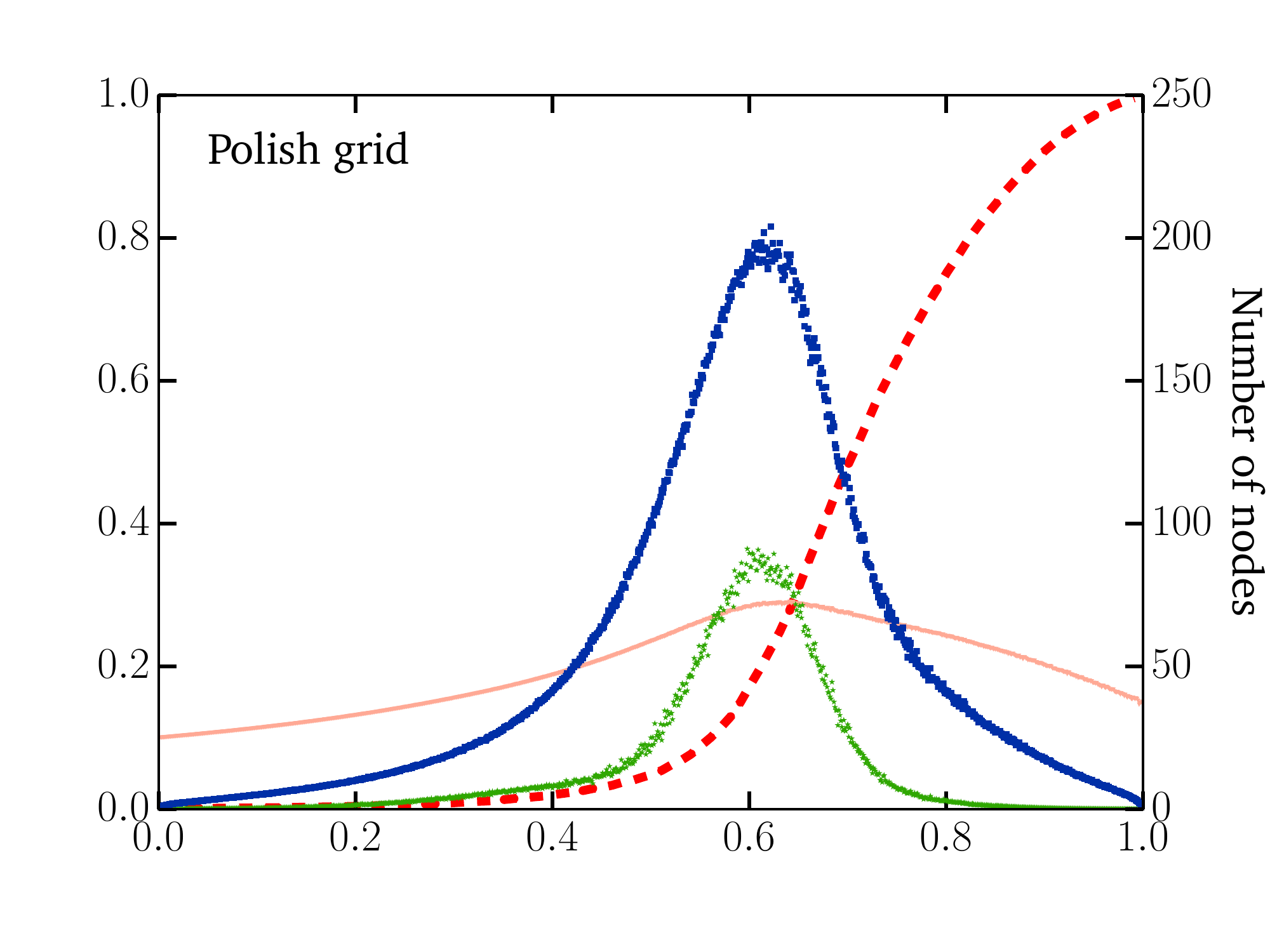}
\includegraphics[trim={0 0 1cm 1cm},clip,width=0.49\linewidth]{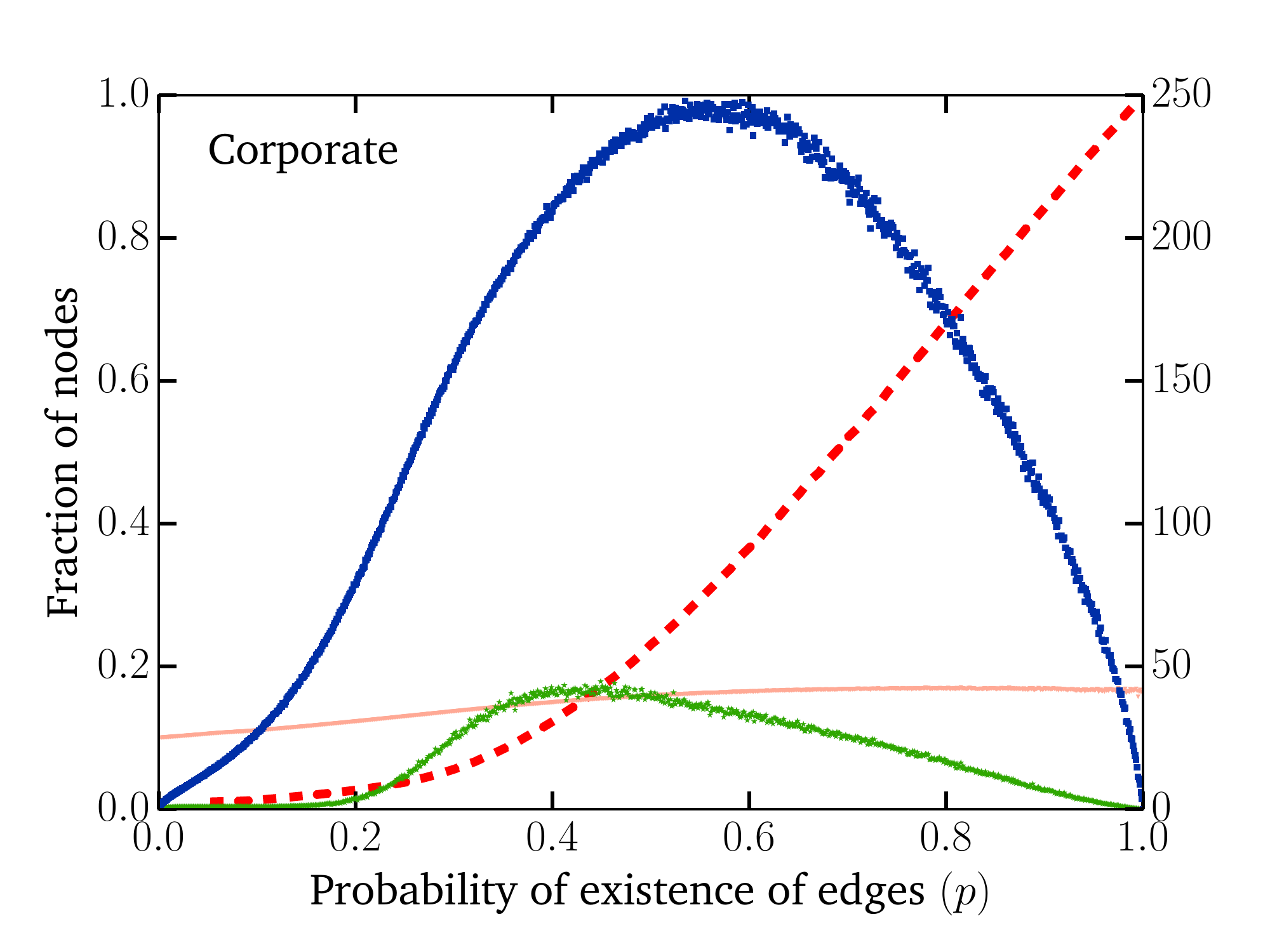}
\includegraphics[trim={1cm 0 0 1cm},clip,width=0.49\linewidth]{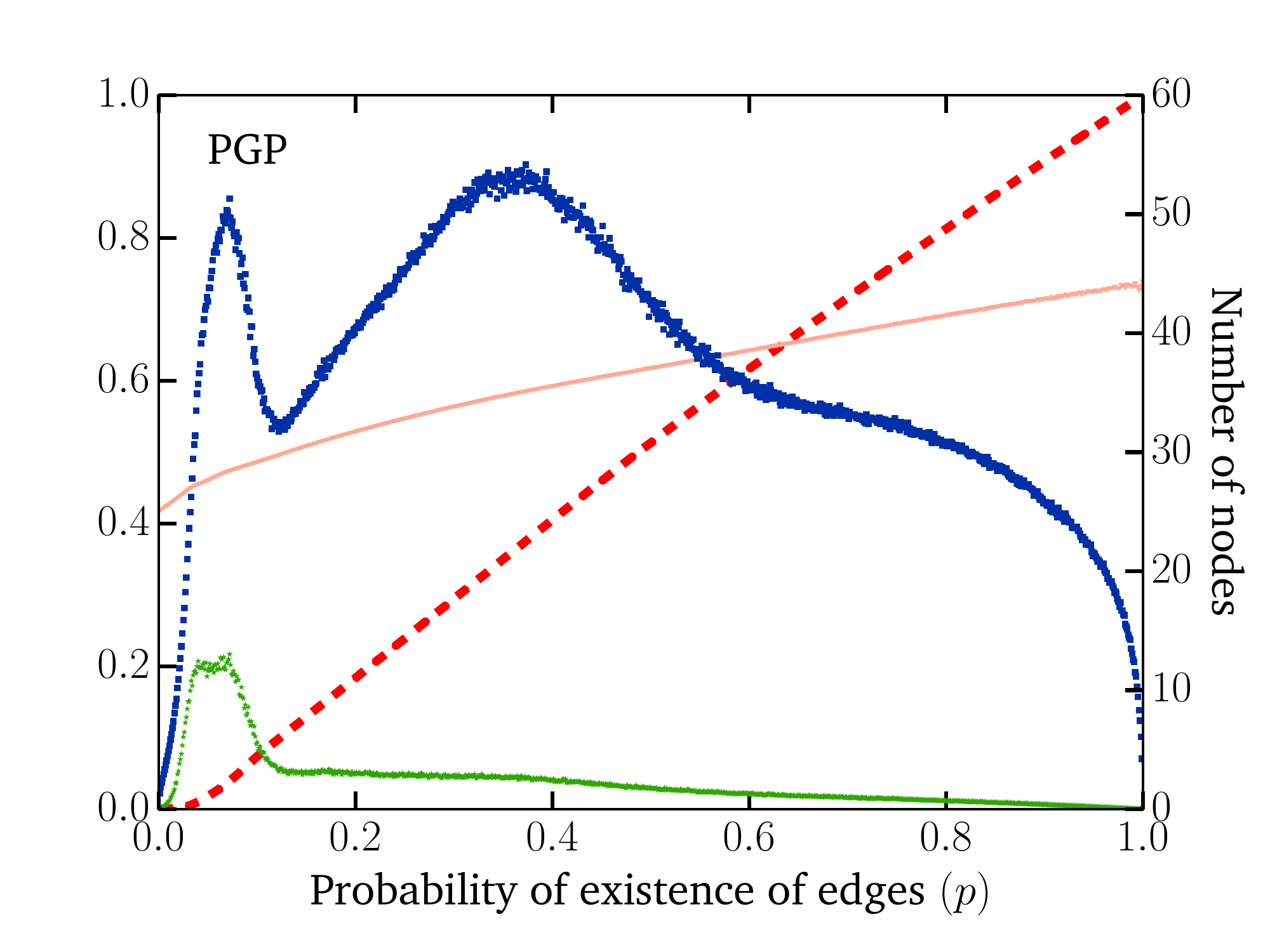}
\caption{\textbf{Detection of the phase transition on four real networks using three common measures.} We study the phase transition as we did in Fig.~\ref{fig:theo_detection}. We now use four real networks: an American power grid \cite{Watts1998}, a Polish power grid \cite{zimmerman2011matpower}, a social network among boards of directors of public Norvegian companies \cite{Seierstad2011} and the web of trust of the PGP encryption algorithm \cite{Boguna2004}. These networks were chosen to highlight the common problems with standard methods to detect the phase transition in real complex systems: (i) They do not necessarily agree, (ii) they can peak once the order parameter is already very large, and (iii) they can peak more than once.}
\label{fig:detection}
\end{figure*}

This paper studies our ability to detect and characterize the percolation phase transition as follows. In Sec.~\ref{sec:num}, we outline and test existing methods to numerically detect phase transitions in a variety of real complex networks. In Sec.~\ref{sec:smeared}, we interpret the results of the previous section using the perspective of smeared phase transitions. We show how the phase transition is generally not a clean transition, but rather a sum of sequential phase transitions within inhomogeneities such as modules, core-periphery structures or degree classes. In Sec.~\ref{sec:anal}, we briefly discuss our results on finite size effects under the lens of message passing and other state-of-the-art analytical methods. We also propose a measure of local susceptibility to potentially identify smeared transitions.
%
%
%
%
%
\section{Phase transition detection\label{sec:num}}
%
\begin{figure*}
\centering
\includegraphics[width=\linewidth]{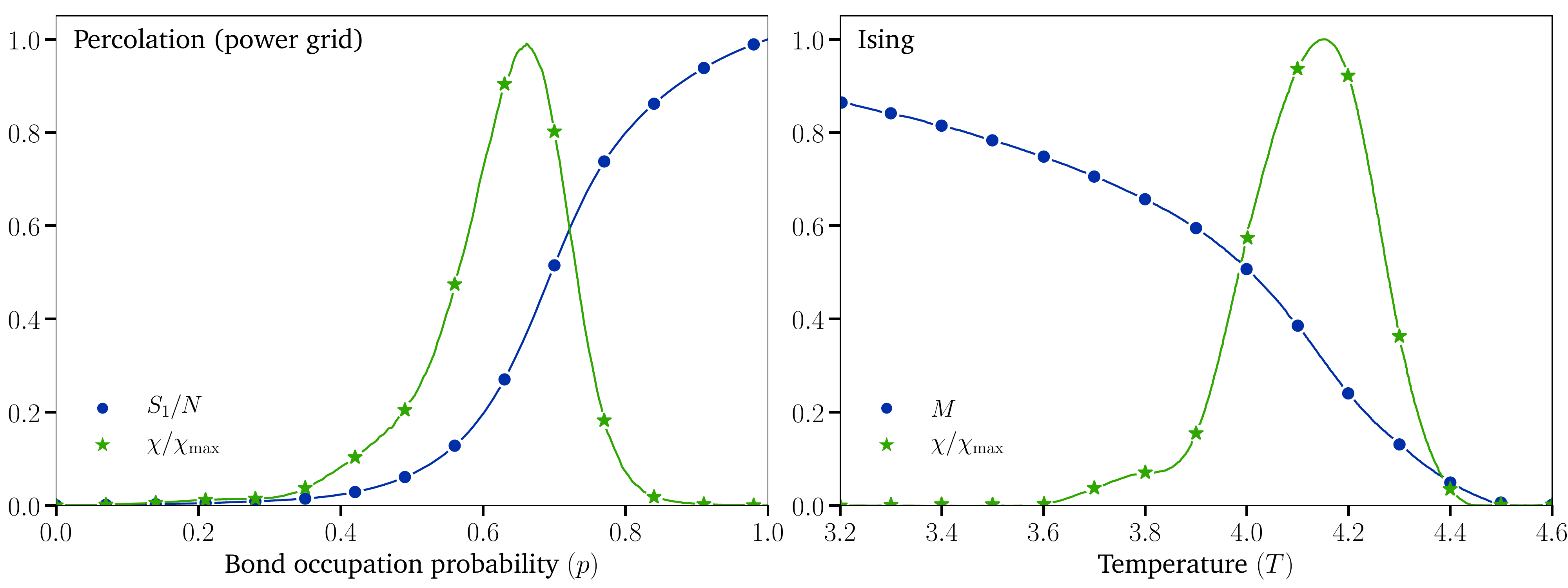}
\caption{\textbf{Smeared phase transitions in two different models.} (a) Simulations of percolation on a power grid of size $N=4941$ where the relative size of the LCC is followed as the external parameter $p$ (bond occupation probability) is varied. (b) Simulations of a three-dimensional Ising model in a cubic lattice of size $N=20000$ with planar defects where the average magnetization is followed as the external parameter $T$ (temperature) is varied, reproduced from Ref. \cite{sknepnek2004smeared}. In both cases, we attempt to detect the phase transition using the susceptibility of the system. Unfortunately, in both cases, susceptibility peaks when the order parameter is roughly equal to a third of its maximal value.}
\label{fig:smeared}
\end{figure*}

To numerically investigate percolation on real networks, we have to adapt the theoretical definition of the order parameter. The relative size of the GCC can be approximated in practice by the ratio $S_1/N$, where $S_1$ is the size of the largest connected component (LCC) and $N$ is the number of nodes in the network (i.e., system size). Moreover, to avoid confusing sub-critical but large components with a super-critical component, we consider $S_1$ as non-extensive whenever it is smaller than $0.01N$.

We consider three methods to detect the position of the percolation threshold which are based on two quantities that are known to peak in homogeneous phase transitions (see Fig.~\ref{fig:theo_detection}): The susceptibility and the average size of the small, or non-extensive, components. The susceptibility measures the expected response of the system as the fraction $p$ of occupied edges (or nodes) is varied. Since the derivative of the order parameter is discontinuous at the phase transition, the susceptibility diverges. These three methods to detect the percolation transition are:
\begin{itemize}
  \item Method \#1: We denote $S_{1,i}$ the size of the LCC in the $i$-th simulation of percolation such that $S_1$ is the average of $S_{1,i}$ over all simulations $i$. The susceptibility $\chi_1$ of $S_1$ can be written as:
  \begin{equation}
    \chi_1 = \frac{\sum _i \left(S_{1,i} - S_1\right)^2}{\sum_i S_{1,i}} \; .
  \end{equation}
  As per classic percolation theory, $\chi_1$ peaks, or diverges in the limit of infinite system size, at the phase transition \cite{Colomer-de-Simon2014,radicchi2015predicting}. 

  \item Method \#2: Before the phase transition, the expected size of small components $\langle s \rangle$ should increase monotonously with $p$ until it grows very large (or, again, diverges in the limit of infinite system size) and form the GCC. After the phase transition, the largest small components are increasingly absorbed by the GCC such that the average size of the small components decreases and typically remains of the order of a handful nodes. One can therefore look for a peak in the average size of small components --- all components other than the LCC --- to identify the phase transition \cite{Allard2017a}.

  \item Method \#3: The third method is a refinement of the previous one and only looks at the size $S_2$ of the second largest connected components (SLCC) \cite{zhang2017spectral}. It typically leads to a refined and more evident peak because of its greater amplitude (i.e., $S_2 \geq \langle s \rangle$) and because of its sharper decrease since the SLCC is typically quickly absorbed by the LCC.

\end{itemize}

We apply these methods to four real networks and Fig.~\ref{fig:detection} shows that they do not identify the same clean phase transition but rather behave in very unexpected ways. First and foremost, they do not always agree nor do they always rank consistently or provide bounds on what we would consider the actual phase transition. Second, the peak is rarely sharp, with the width at half-amplitude often spanning over 10\% of the parameter space. Third, and most surprisingly, Method \#2 does not always peak and some methods can peak more than once. 

\begin{figure*}
\centering
\includegraphics[width=0.45\linewidth]{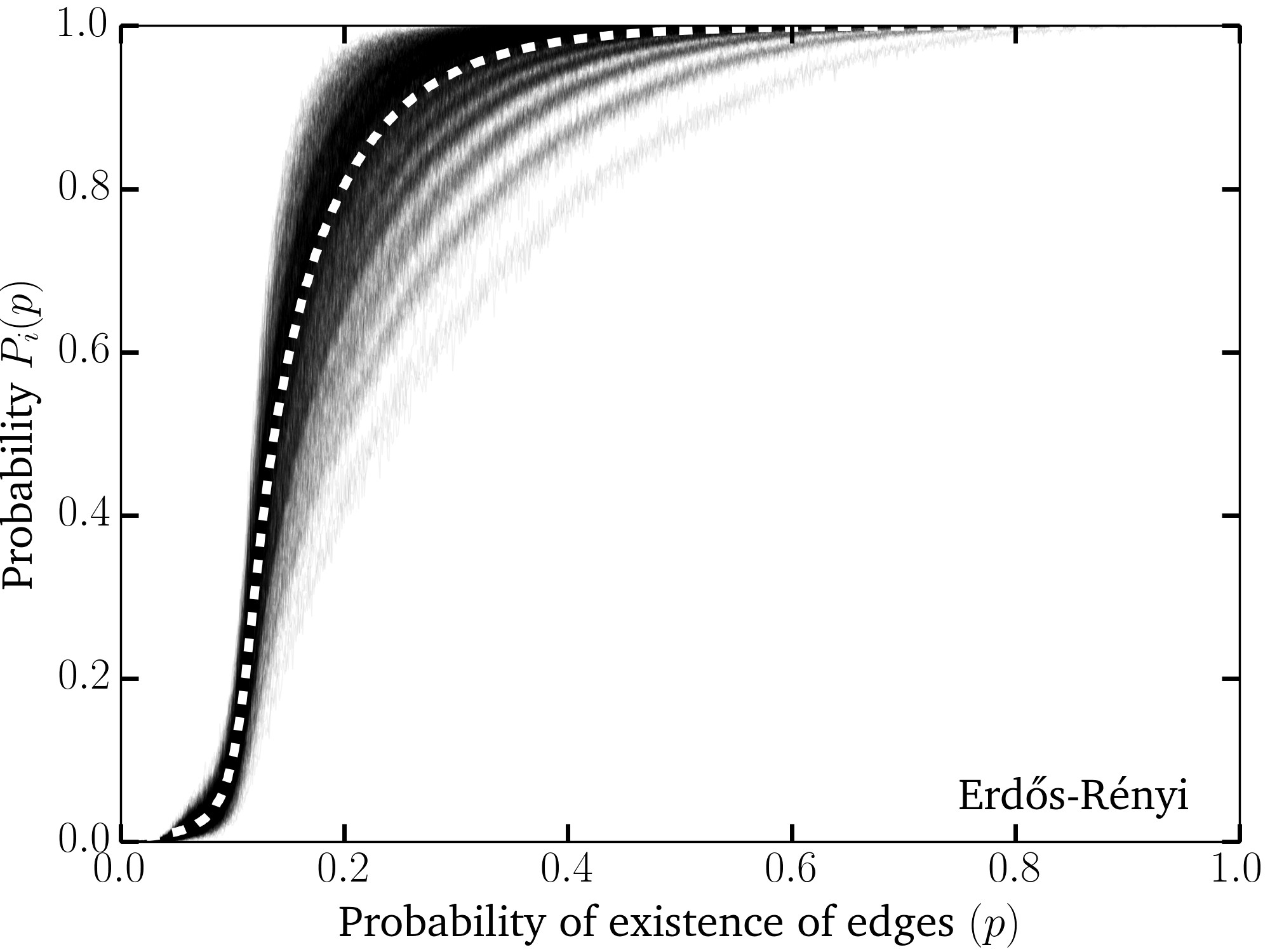}
\includegraphics[trim={0.75cm 0 0 0cm},clip,width=0.433\linewidth]{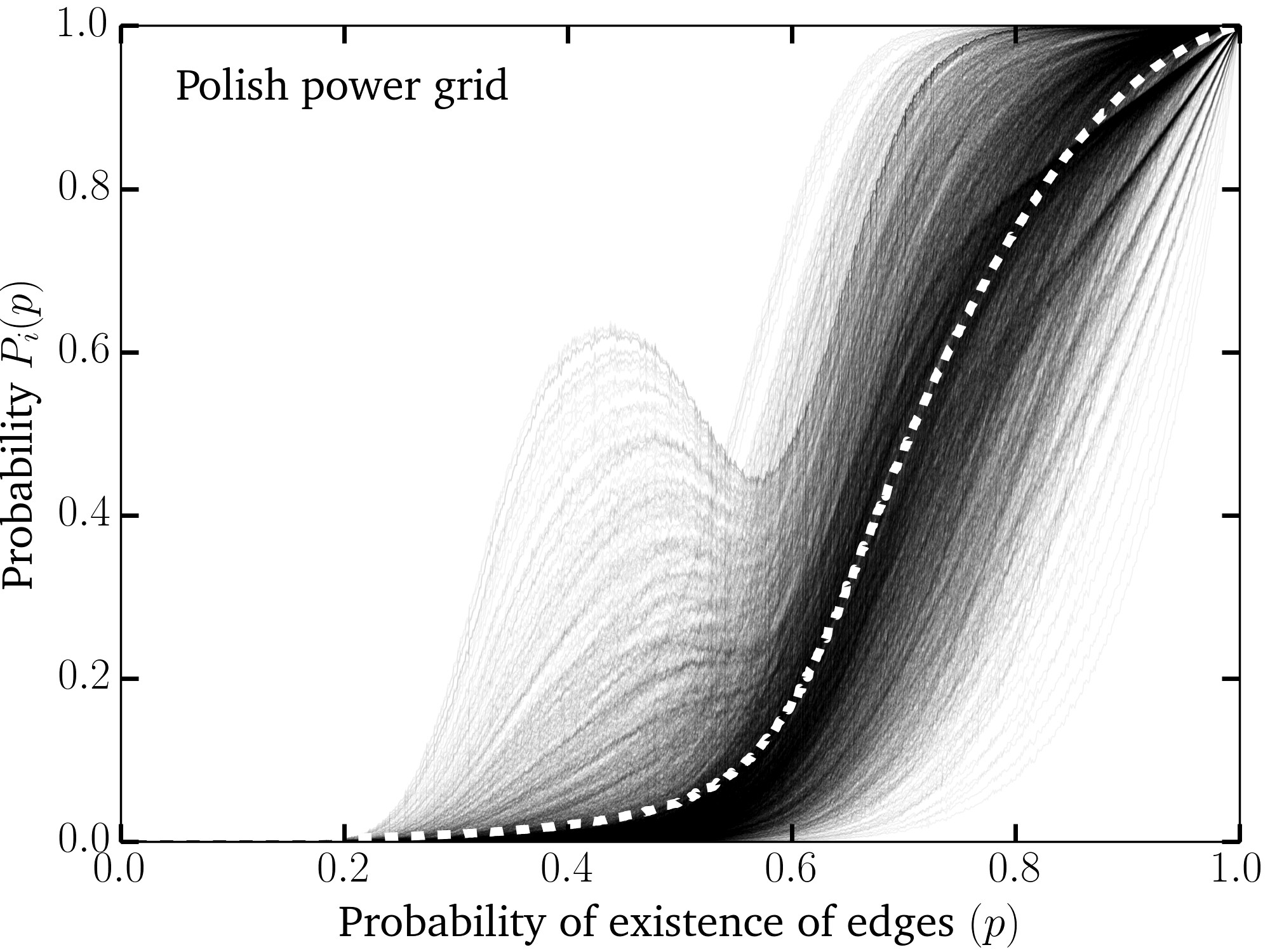}
\caption{\textbf{Local order parameter.} (left) Results of percolation simulations on a Erd\H{o}s-R\'enyi (ER) random graph of size 1000 with average degree 10. We show the global order parameter (i.e., the relative size of the LCC) in white, while the underlying gray curves show the probability $P_i(p)$ that every individual node $i$ is found in the LCC. Despite the small size of the system, we find a relatively clean phase transition; meaning that the global order parameter accurately describes the behavior of observed around individual nodes. Most importantly, all curves of the local order parameter are the steepest at the same point (i.e. the maxima of $dP_i(p)/dp$ correspond to the phase transition). (right) Results of percolation simulations on a Polish power grid shown with the same color scheme. We find that while the global order parameter represents the average of all $P_i(p)$ (by definition), it is not representative of the behavior of every $P_i(p)$.}
\label{fig:polishtraj}
\end{figure*}

%
%
%
%
%
\section{Smeared phase transitions\label{sec:smeared}}
%
\begin{figure*}
\centering
\includegraphics[trim={0.0cm 0.8cm 0 0cm},clip,width=0.45\linewidth]{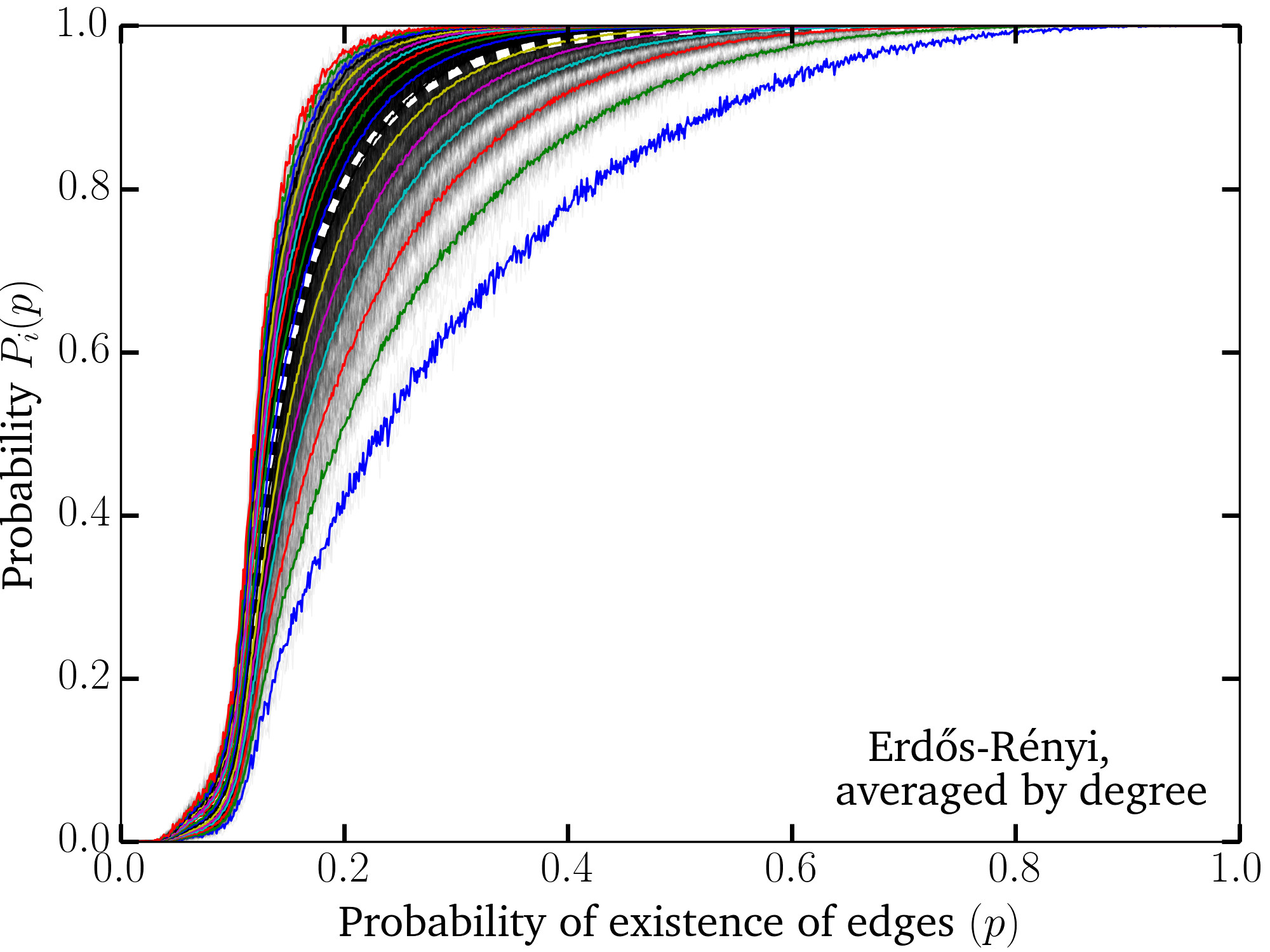}
\includegraphics[trim={0.75cm 0.8cm 0 0cm},clip,width=0.433\linewidth]{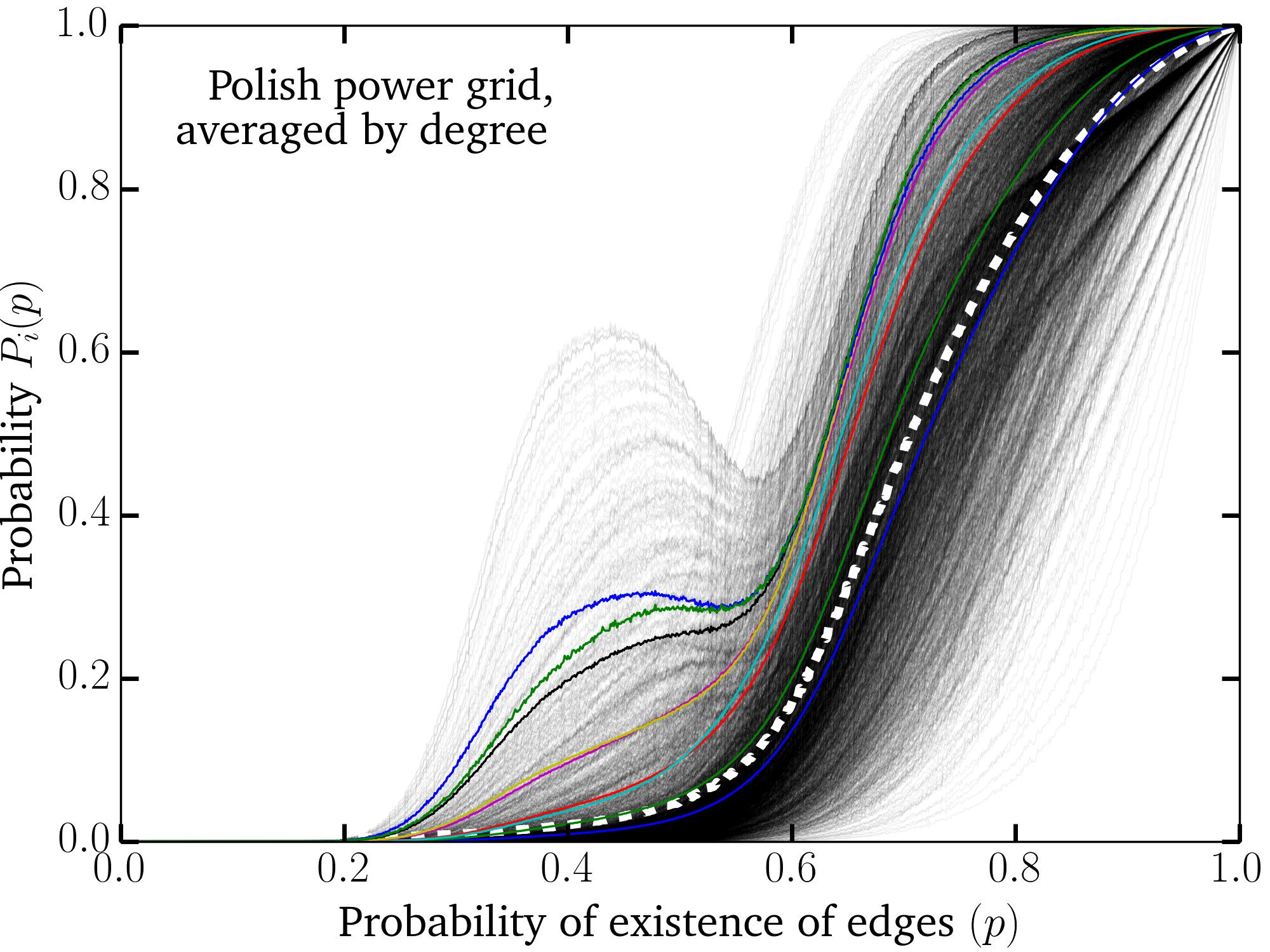}\\
\includegraphics[width=0.45\linewidth]{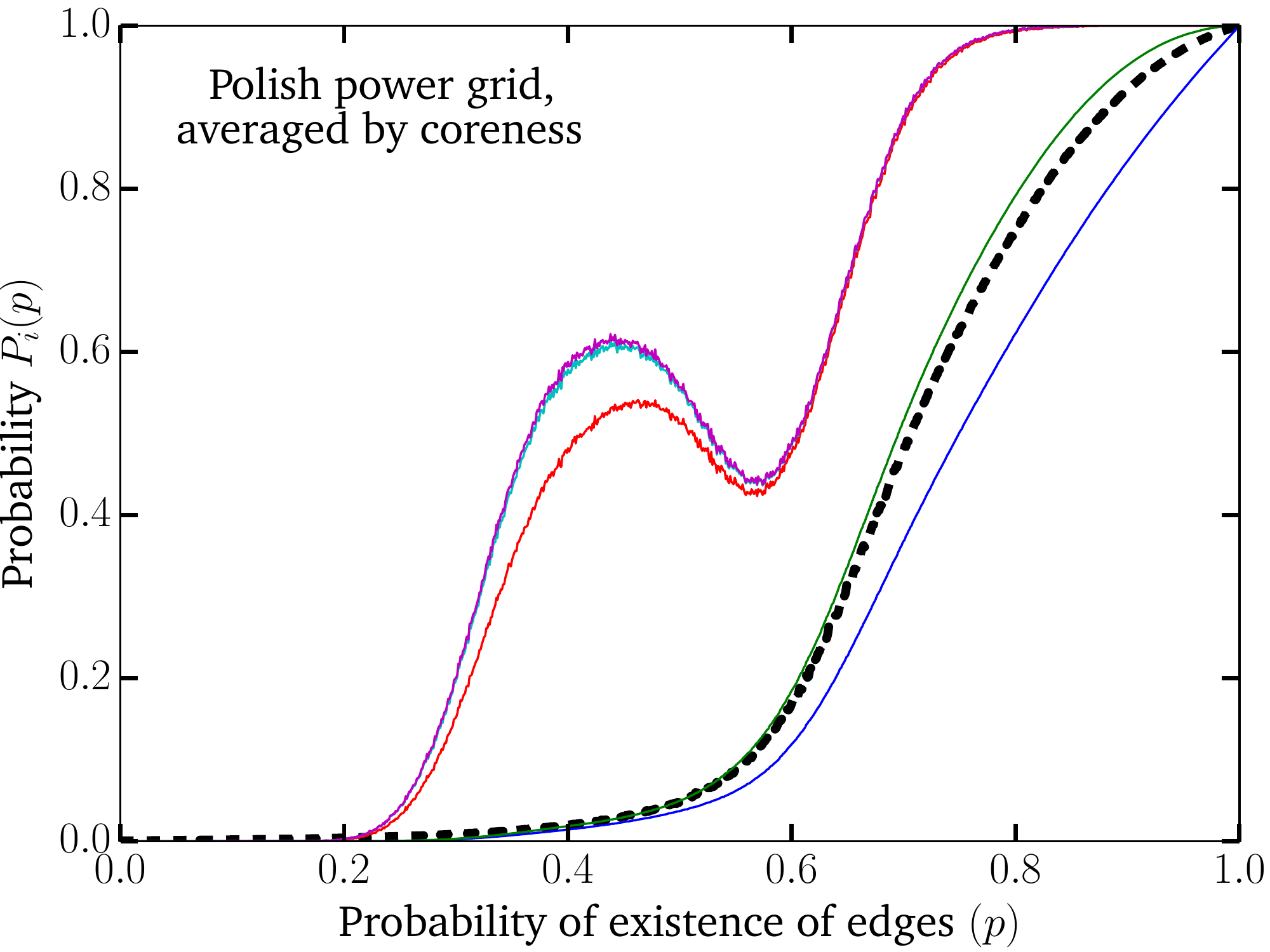}
\includegraphics[trim={0.75cm 0 0 0cm},clip,width=0.433\linewidth]{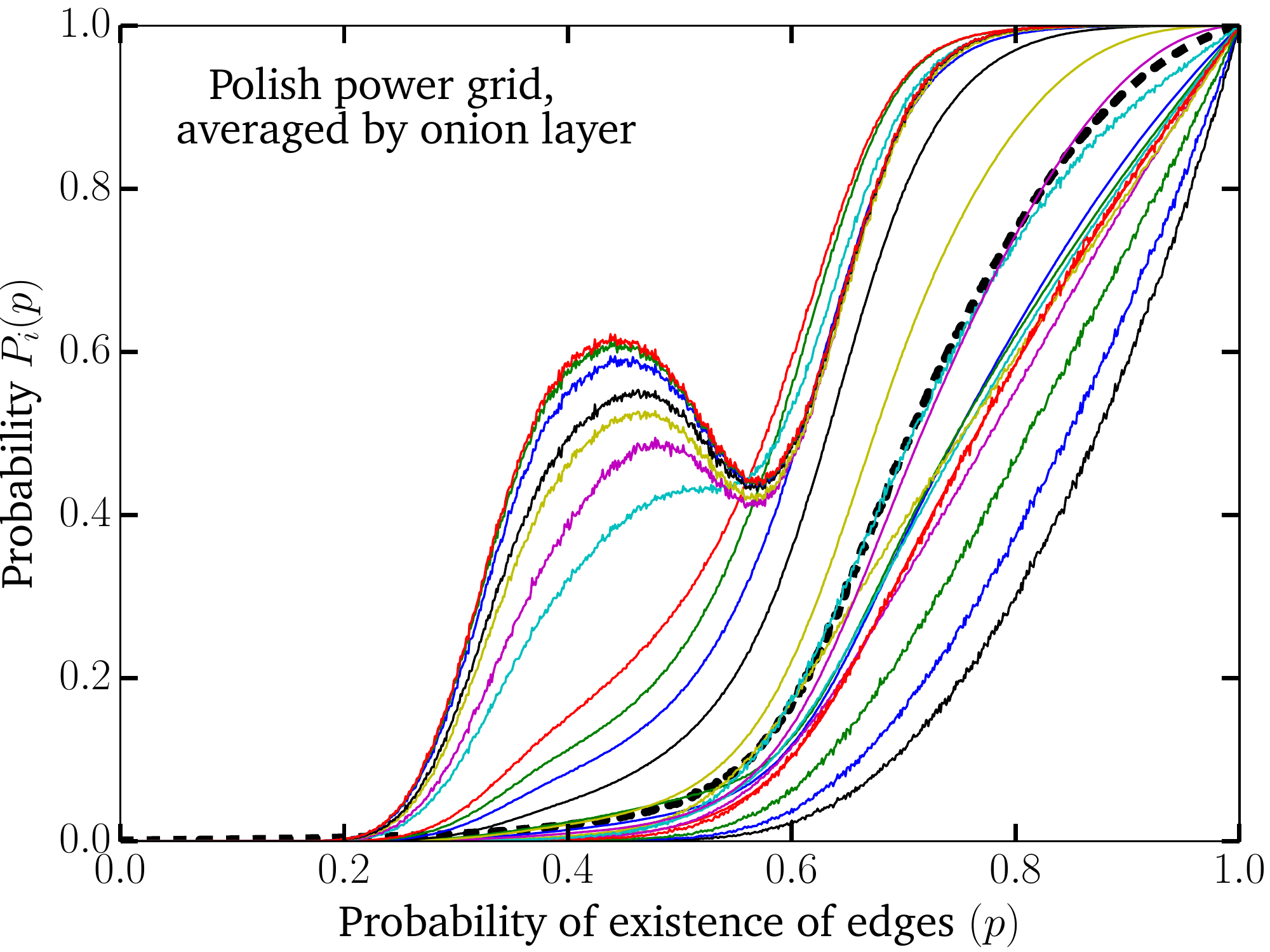}
\caption{\textbf{Order parameter in different centrality classes.} (top row) We compare the curves shown in Fig.~\ref{fig:polishtraj} to averages (additional colored curves) based on the degree of nodes. The new curves show the average probability that a particular node is found in the LCC based on its degree (i.e. from the bottom up, we plot averages over degree 1, 2, 3 and so on). (bottom row) The color curves represent the average of $P_i(t)$ over nodes $i$ belonging to given (left) $k$-shells of the $k$-core decomposition and (right) layers of the onion decomposition. These more complicated centrality metrics capture progressively more topological diversity, especially around the most and least central nodes.}
\label{fig:centrality}
\end{figure*}

The two first problems observed at the end of the previous section are related --- the peaks of the observables do not align and are not sharp --- and correspond to what we would expect given strong finite size effects. Indeed, the finite size of real systems tends to smooth out phase transitions of all nature. In the case of percolation, this happens because large but non-extensive components are hard to distinguish from a GCC if we cannot change the system size to investigate scaling relations. Just below the threshold, small components larger than our LCC criteria ($1\%$ of system size) exist such that a non-zero order parameter below $p_c$ is numerically observed. These effects are inherent of the use of the phase transition framework to real finite size systems, which in theory only applies in the infinite size limit. And while there are methods available to account for some finite size effects \cite{bianconi2018rare} and other rare fluctuations \cite{bianconi2017fluctuations}, the basic phenomenology of the transition, averaged over all possible realizations, always remains the same.

Finite size effects therefore fall short of explaining the enormous width of the susceptibility peaks in Fig.~\ref{fig:detection}, nor can they explain the disagreement between the three measures or why they peak more than once.

\subsection{Empirical results}

Another possible explanation is that we are dealing with smeared phase transitions. One classic example is that of the Ising model in systems with defects \cite{sknepnek2004smeared}. The Ising model considers spins, which can take a $+1$ or $-1$ value, laying on the nodes of a regular lattice. At high temperature, the spins are independent of each other and free to take either value such that the global (or average) magnetization of the system is zero; this is the disordered or paramagnetic phase. As temperature decreases, the interactions force correlations and the system eventually enters a correlated state with non-zero global magnetization; this is the ordered or ferromagnetic phase. Because of the regular structure of lattices, and because all interactions are equal, the system is highly homogeneous. This homogeneity leads to a clean phase transition: There is no spatial variation in thermodynamic observables and all mesoscopic domains undergo an identical phase transition. This also means that in the thermodynamic limit, we see a vanishing width or variance in the distribution of the order parameter across domains. This property is called self-averaging.

One of the most powerful aspect of phase transitions is their resilience to microscopic details in the structure or rules of our models. For example, we can introduce significant defects in the lattice on which the Ising model occurs without destroying the sharpness of its phase transitions. Defects such as weakening/strengthening or removing/adding edges do not affect the phenomenology of the model as long as they are not strongly correlated (e.g., random micro- or mesoscopic noise in space).

However, the observed phase transition can change drastically when strongly correlated defects are introduced. A classic example is that of the Ising model in a three-dimensional cubic lattice with planar defects creating weaker bonds. The phase transition observed in that model is compared to percolation on a power grid on Fig.~\ref{fig:smeared}, and the phenomenological similarity is striking. This smearing is related, at least physically, to the Griffiths phenomenon \cite{munoz2010griffiths} but different through the long-range order established by correlated defects \cite{sknepnek2004smeared}.

If we accept that our inability to accurately locate the percolation phase transition in real complex networks are not solely due to finite effects but potentially also to correlated defects, the question becomes: What is the source of this disorder? First, since complex networks are not regular systems like lattices, ``disorder'' is the norm and we instead look for correlated \textit{inhomogeneities}. Second, we can detect these inhomogeneities by using the definition of the smeared phase transition. We thus look for subsets of nodes around which the local order parameter deviates from the global order parameter. More specifically, we wish to identify sets of nodes $\{i\}$ such that the probability $P_i(p)$ that node $i$ is in the LCC under occupation probability $p$ is not well approximated by $S_1(p)/N$.

In Fig.~\ref{fig:polishtraj}, we show the curves $P_i(p)$ for all nodes $i$ in a homogeneous random network and in a Polish power grid network, and compare them to the corresponding average $\langle P_i(p) \rangle = S_1(p)/N$. The main conclusion that can be drawn from this experiment is that we can expect significant variance in the distribution of the order parameter in both random and real complex networks, but significantly moreso in the latter. Without an idea of how the variance would scale with system size, it is impossible to rule out the possibility of a finite size effect, but it does seem to be more in line with the smeared phase transition interpretation. We can investigate the nature of these inhomogeneities by averaging $P_i(p)$ over subset of nodes $i$ who share a given structural property. 

In Fig.~\ref{fig:centrality}, the curves $P_i(p)$ are compared to the averages obtained over different centrality classes: Degree centrality given by the number of edges on a given node, $k$-core centrality given by the largest $k$ such that a node is in the maximal subset of nodes with degree at least $k$ among one another \cite{batagelj2011fast}, and the onion decomposition which assigns a layer centrality to every node based on when they are removed in the $k$-core decomposition \cite{Hebert-Dufresne2016a}. Degree heterogeneity is sufficient to capture all the inhomogeneities observed in the Erd\H{o}s-R\'enyi random graph but not in the power grid. However, using the onion decomposition, we now capture much more directly the fact that the system is divided into two (or more) subsystems composed of nodes with very different centrality, not necessarily related to degree but to their position in the network structure.

\subsection{Synthetic results}
\begin{figure*}
\centering
\includegraphics[width=0.49\linewidth]{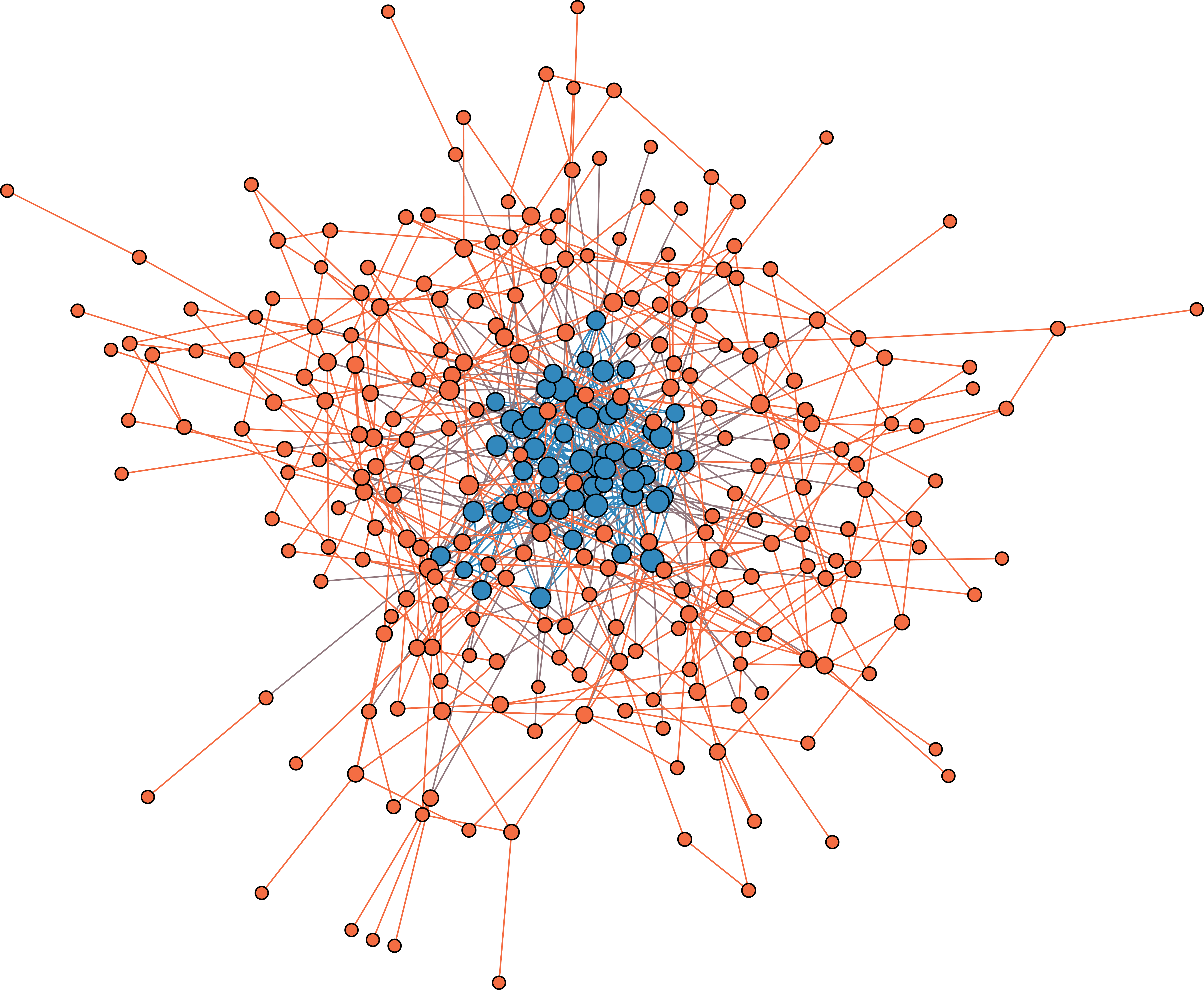}
\includegraphics[width=0.49\linewidth]{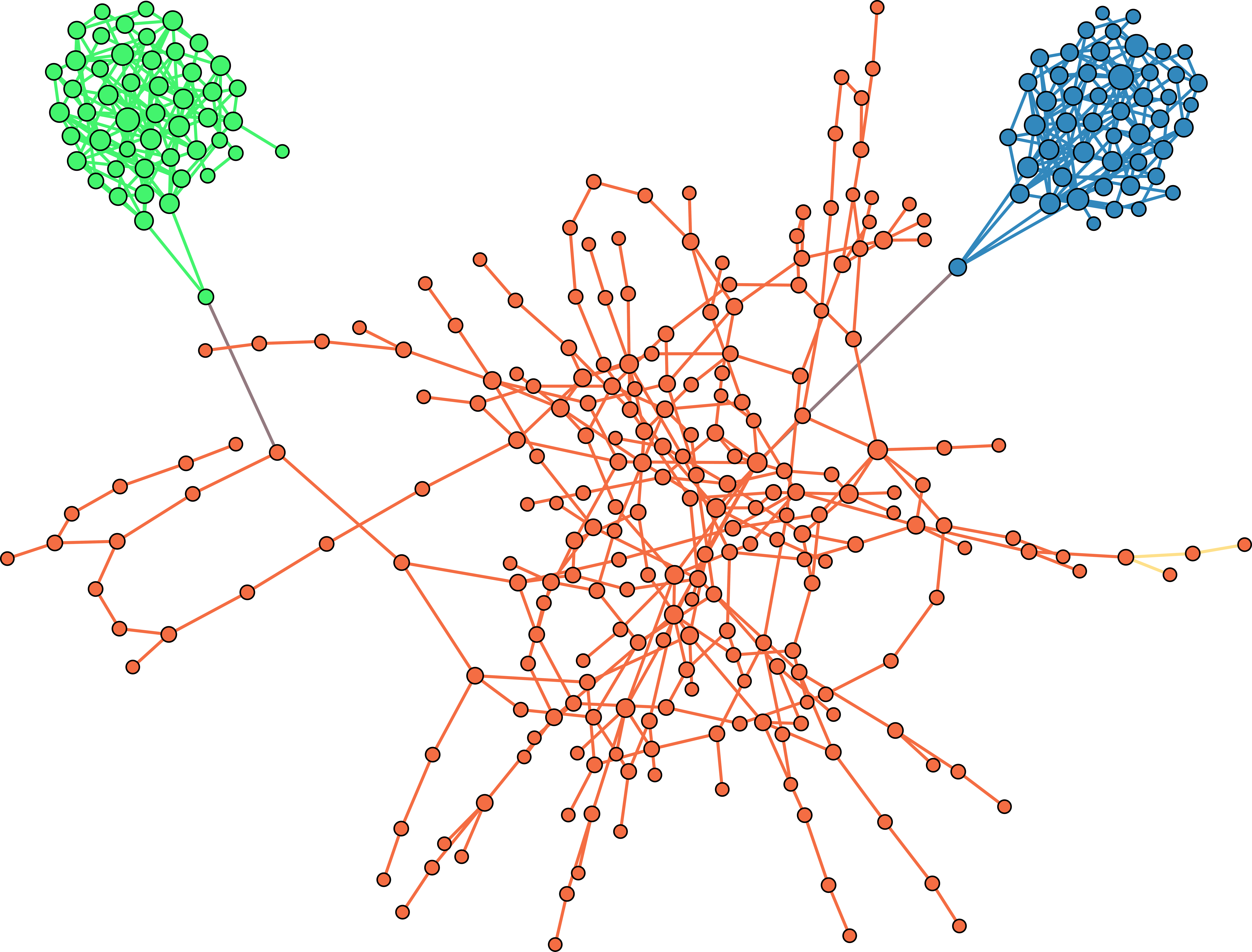}\\
\includegraphics[trim={0 -2cm 0 0},clip,width=0.49\linewidth]{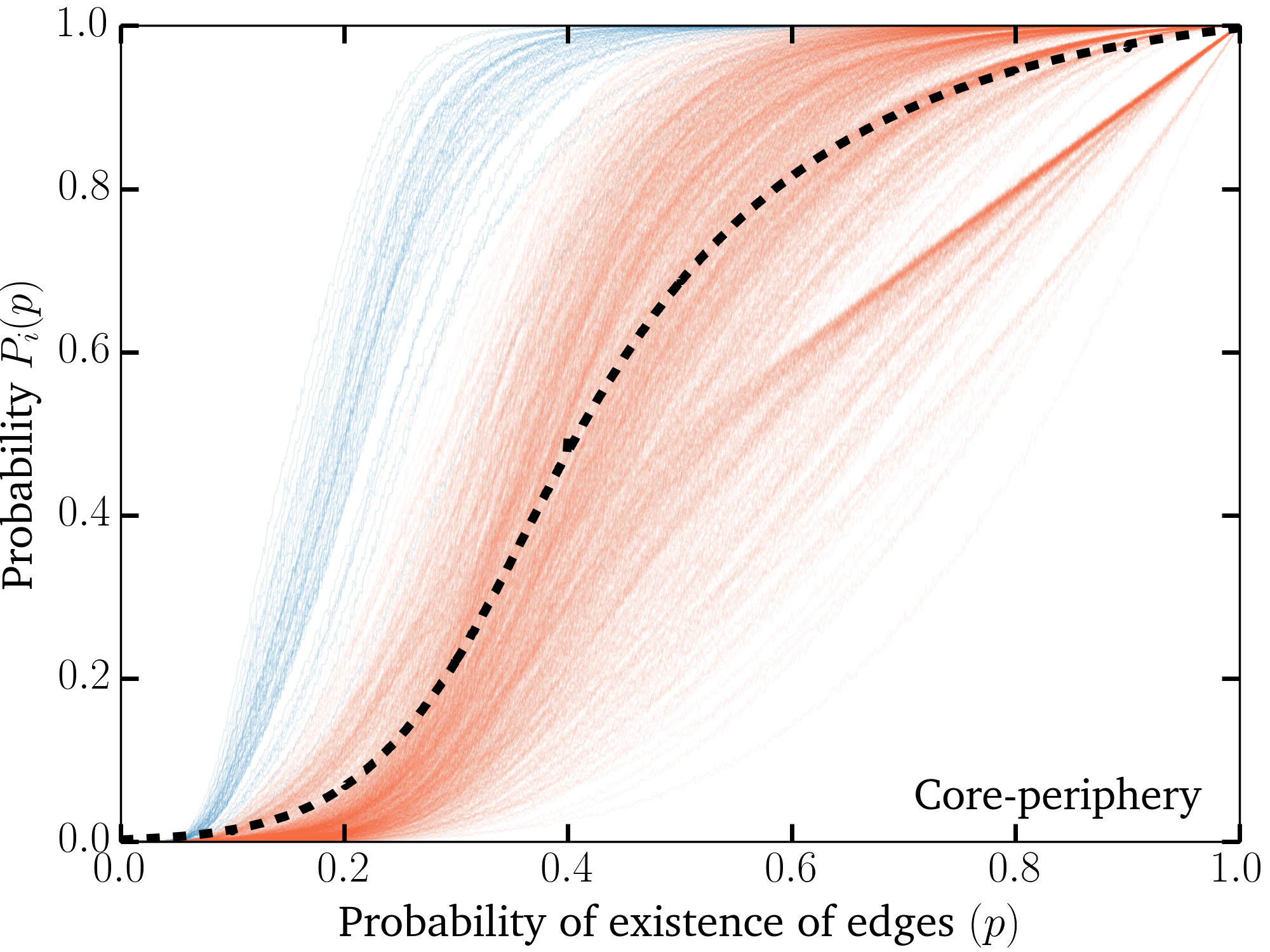}
\includegraphics[trim={0.75cm -2cm 0 0},clip,width=0.472\linewidth]{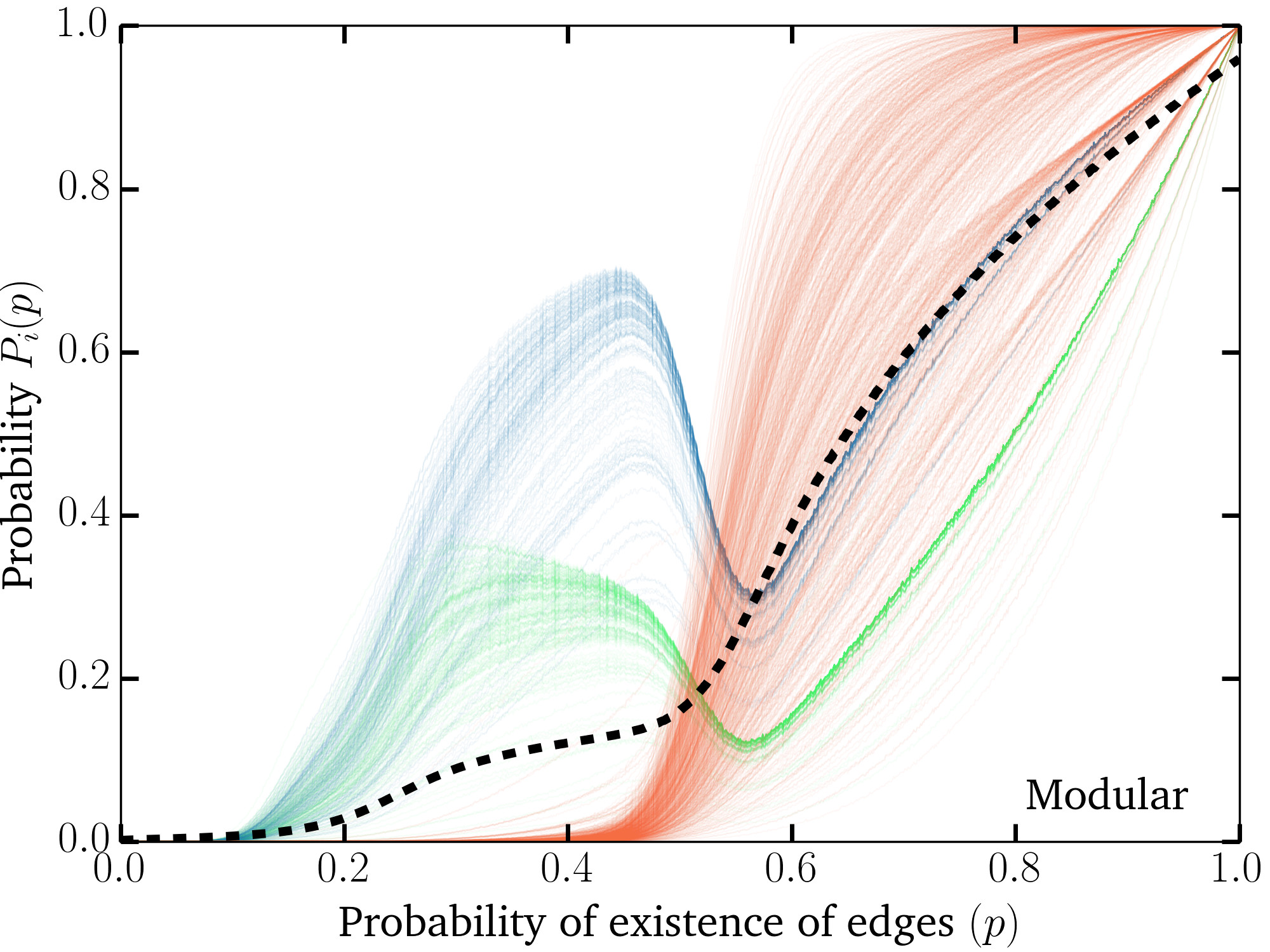}
\caption{\textbf{Local order parameter in two toy networks with core-periphery and modular structure.} (left column) Core-periphery: The core is an Erd\H{o}s-R\'enyi (ER) graph of $N_1$ nodes with density $\rho_1$ and the periphery is an ER graph with $N_2>N_1$ nodes are connected to each other and to the core with density $\rho_2 < \rho_1$. (right column) Modular structure: There are three distinct ER graphs of size $N_1$, $N_2$ and $N_3$ with densities $\rho _1$, $\rho _2$ and $\rho _3$ respectively. $N_1$ and $N_2$ are both connected to $N_3$ with a single random edge. (top row) We show examples of the structures considered in this experiment. (bottom row) The curves $P_i(p)$ in a core-periphery produced with $N_1 = 51$ and $N_2 = 1001$ with densities $\rho_1 = 0.15$ and $\rho _2 = 0.003$, and in a modular structure with $N_1 = N_2 = 201$ and $N_3 = 1101$ with densities $\rho_1 = \rho_2 = 0.025$ and $\rho _3 = 0.002$.}
\label{fig:toy}
\end{figure*}

\begin{figure*}
\centering
\includegraphics[width=0.49\linewidth]{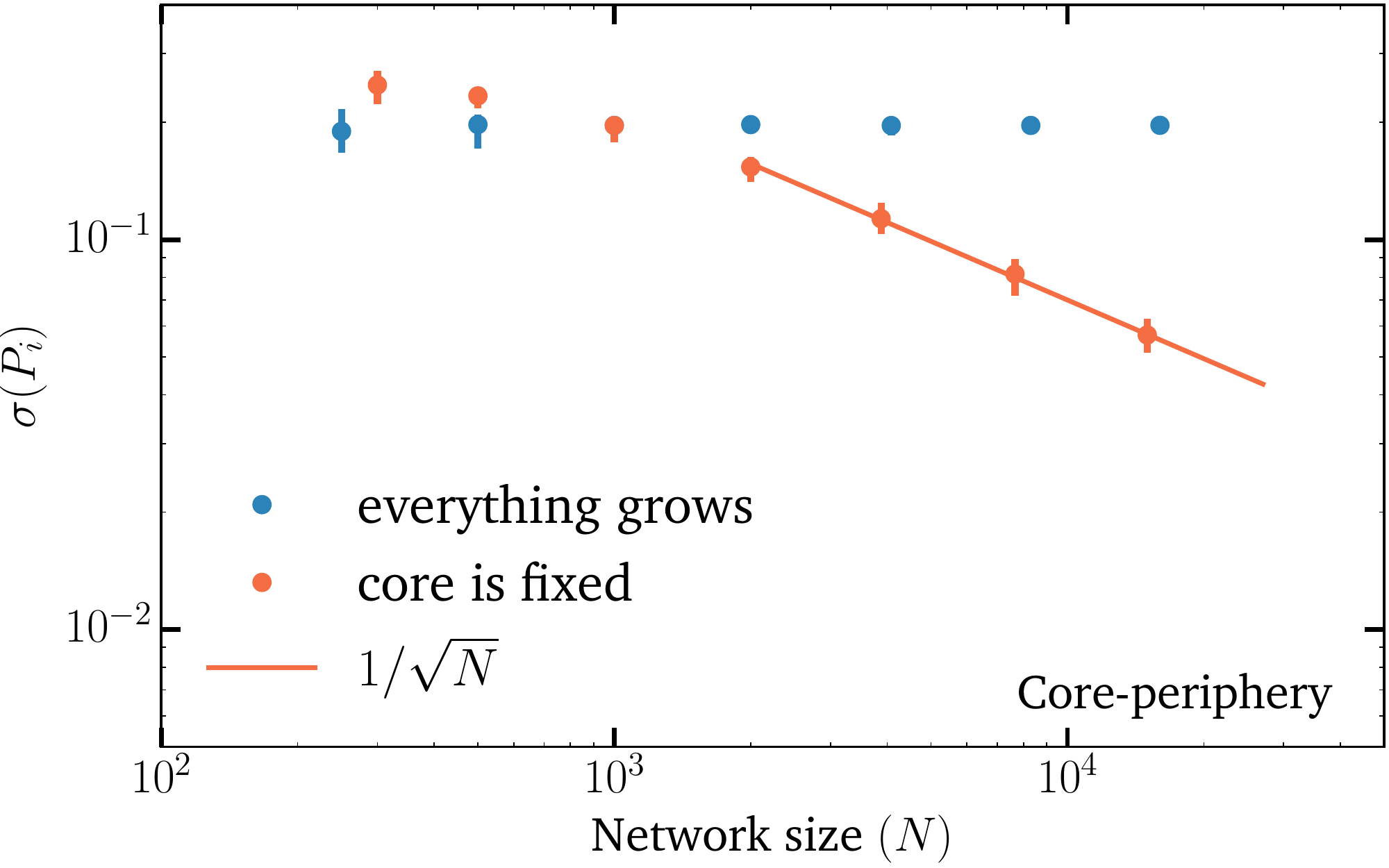}
\includegraphics[trim={0.75cm 0 0 0},clip,width=0.472\linewidth]{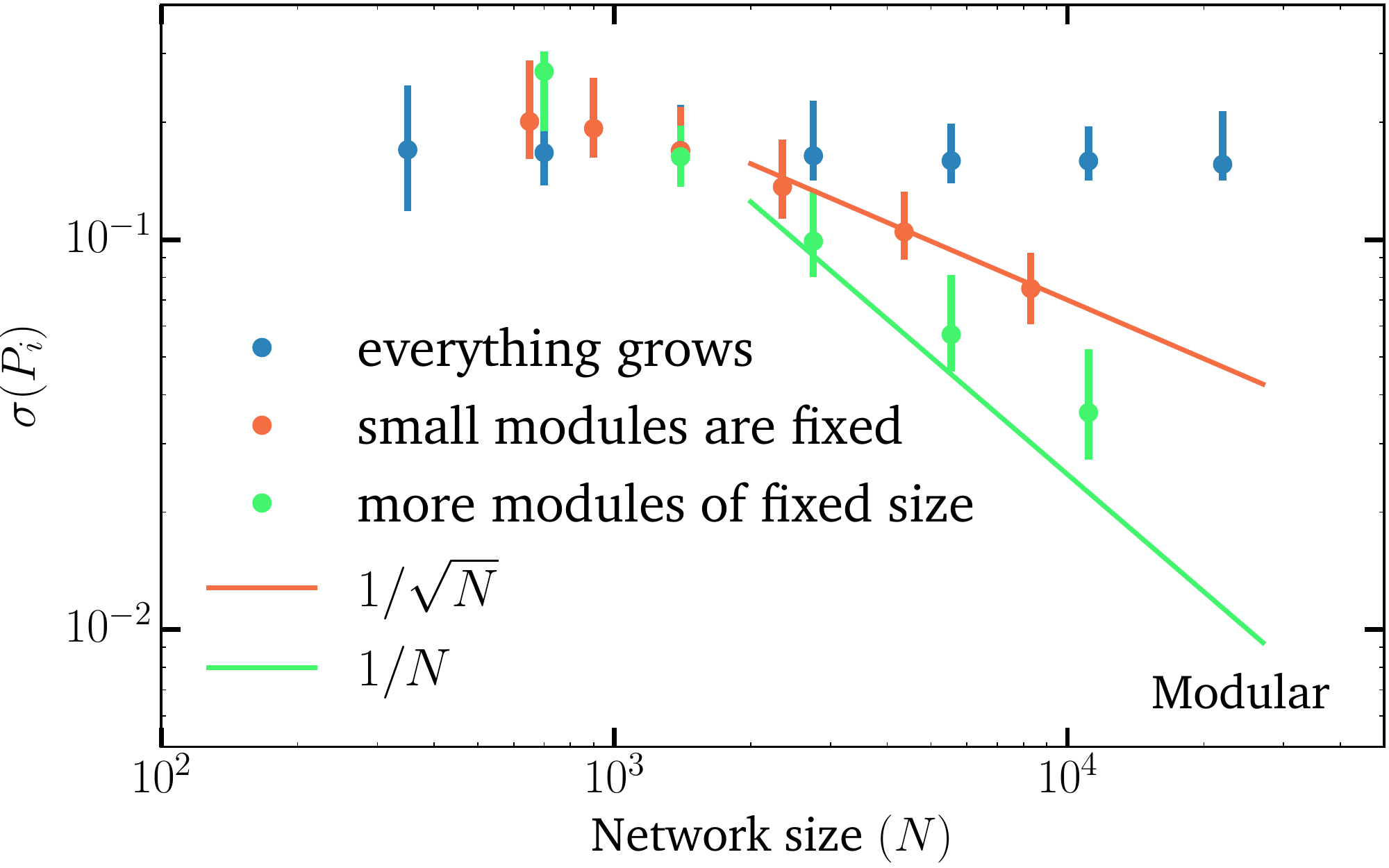}
\caption{\textbf{Vanishing and finite-randomness in two toy networks with core-periphery and modular structure.} We use the core-periphery (left) and modular (right) structures presented in Fig.~\ref{fig:toy}. We follow the standard deviation of $P_i(p)$ at $p=0.3$ when the network size is increased while keeping the average degree of subsystems fixed by varying their densities. In the core-periphery, growing a network means either that the size of both the core and the periphery grow together (option 1, in blue), or that only the periphery grows and the size of the core remains fixed (option 2, in orange). In the modular structure, growing a network means, again, growing all modules together (option 1, in blue), growing only the largest module with fixed number and size for small modules (option 2, in orange), or growing the largest module while adding smaller modules of fixed size (option 3, in green). The markers correspond to the average and error bars to the full range of observed values. Lines showing theoretical bounds of $N^{-1/2}$ and $N^{-1}$ have been added to guide the eye. The $N^{-1/2}$ behaviour corresponds to typical vanishing finite size effects. The faster decay in $N^{-1}$ corresponds to competing finite size effects, obtained for example when adding small modules makes it less likely for any given module to contain the LCC.}
\label{fig:toy2}
\end{figure*}

To investigate the results of mesoscopic structure on the percolation phase transition, we study the set of probabilities $P_i(p)$ introduced in the previous section within 2 toy networks. Both are based on the traditional Erd\H{o}s-R\'enyi (ER) random graph investigated in Fig.~\ref{fig:polishtraj}, a simple network of $N$ nodes where every unique pair of nodes is independently connected with probability $\rho$. Our first toy model produces a core-periphery structure using two nested ER graphs: An inner smaller graph of size $N_1$ with density $\rho_1$ and a larger outer graph of size $N_2 > N_1$ where nodes are connected among themselves and to the inner graph with density $\rho_2 < \rho_1$. The second toy network is produced by a set of independent ER graphs of different size and/or density which are then connected by a single edge; this results in networks with a strong modular structure. The top row of Fig.~\ref{fig:toy} provies a typical realization of these two toy models.

The bottom row of Fig.~\ref{fig:toy} shows the $P_i(p)$ curves obtained with the two toy models alongside the average value over all nodes. In the case of the core-periphery structure, the percolation process nucleates in the core at an occupation probability $p$ much lower than the known threshold $1/3$ of the periphery. Peripheral nodes are still ``activated'' below their threshold due to subcritical spillover from the core into the periphery. Eventually, the periphery activates and creates a sudden rise in the $P_i(p)$ curves of peripheral nodes. Importantly, we find that all curves increase monotonously, as the percolation spreads progressively from the core outward as $p$ increases. 

In the modular networks, the percolation process again nucleates in the denser modules, but eventually the $P_i(p)$ curves corresponding to nodes in denser modules decrease. This is due to the weak coupling between modules: there exist a regime where large connected components exist in each module but are unlikely to connect and are therefore in competition for the title of \textit{largest} connected component. Hence, whether there is a subset of non-monotonous $P_i(p)$ curves can be used to distinguish core-periphery and modular structures. For example, the set of $P_i(p)$ curves observed with the modular networks are strongly reminiscent of the results on the Polish power grid which, in turn, suggests that their is important modularity in its inner cores structure which lead to non-monotonous $P_i(p)$ curves.

We further test the robustness of these inhomogeneities in the distribution of local order parameter $P_i(p)$ by measuring their standard deviation as we increase system size. The results are shown in Fig.~\ref{fig:toy2}. There are different ways to increase the size of these random networks, and we investigate three options. (i) We increase the size of all subnetworks at the same time. (ii) We only increase the size of the largest subnetwork. (iii) We increase the size of the largest subnetwork, and add more subnetworks of a fixed smaller size. Doing so we find that the smeared transition is only preserved when all subnetworks are scaled simultaneously. Interestingly, adding more inhomogeneities (smaller denser modules) of fixed size collapses the transition even faster than not scaling the inhomogeneities at all. Again, this is due to increased competition between inhomogeneities which might all have large connected components competing for the title of the LCC. Importantly, this result stresses the need for \textit{correlated} inhomogeneities to produce smeared phase transitions.
%
%
%
%
%
\section{Discussions\label{sec:anal}}
%

\subsection{The thermodynamic limit and message passing approaches to percolation}

To sum up the results obtained so far, one can characterize a phase transition by looking at the set of probabilities $P_i(p)$ of finding node $i$ in the LCC at occupation probability $p$. The curves $P_i(p)$ can help identify the mesoscopic organization of the structure at hand. Using two simple toy models, we showed that core-periphery leads to monotonously increasing $P_i(p)$ where the core activates first and gradually invades the periphery. Modular structure, on the other hand, leads to non-monotonous $P_i(p)$ where the LCC appears in a dense cluster first, and eventually jumps to a parser, but larger second cluster. In other words, the strength of the coupling between cores and/or modules distinguishes these different results.

To better understand the role of coupling between substructures, we follow the language of Sknepnek \& Volta and distinguish three types of mesoscopic network inhomogeneities \cite{sknepnek2004smeared}. The first one, called vanishing-randomness, corresponds to where the distribution of local thermodynamic observables collapses in the thermodynamic limit, meaning that the system is self-averaging and does not produce true smeared phase transitions. The second and third types of network inhomogeneities are finite-randomness and infinite-randomness, meaning that the distribution of local thermodynamic observables are \textit{not} self-averaging but instead lead to finite or infinite width in the distributions of local order parameters in the thermodynamic limit. Vanishing- and finite-randomness both appear in Fig.~\ref{fig:toy2} depending on how the network grows when taking the thermodynamic limit.

Since this categorization depends on how we take the thermodynamic limit of a system, there are no way to apply our intuition to real systems. Indeed, there are of course no way to know whether the dense cores observed in power grids should scale with the size of the system or not. If the power grid was twice as large, would we find a core of similar size, or twice as many cores, or a single core twice as large? Distinguishing important mesoscale structures from finite size effects is context-dependent and must be done on a case-by-case basis. What we can say however is that most of current mathematical approaches are making that decision for us.

Current state-of-the-art analytical approaches to percolation are based on the message passing approximation (MPA) \cite{Karrer2014}. The MPA takes the entire network structure as an input, but then ignores loops when solving the percolation process. Because of this approximation, the MPA is effectively solving percolation not on the true network, but on an infinite network where there exist an infinite number of copies of every node \cite{faqeeh2015network}. In practice, this means that any modular structure is mapped to a core-periphery structure. For example, when using the MPA on a network with two modules of size $N$ with uniform degrees $k_1$ and $k_2$ and connected by a single edge, we are solving percolation on a core-periphery structure where the core is a $(k_1-1)$-core, and the periphery a $(k_2-1)$-core, interconnected by a bridge in a fraction $1/N$ of nodes (which corresponds to an infinite number of bridges).

In practice, this means that any core-periphery structure will be captured by the MPA, but that modular structure will be mapped to an equivalent core-periphery structure \cite{Allardforthcoming}. For the local order parameter, this implies that the MPA will always predict monotonously increasing $P_i(p)$ curves. Consequently, while it is tempting to dismiss dense but small substructures, it is important to know that the MPA, and other analytical approaches that account for node centrality \cite{Hebert-Dufresne2013, allard2018percolation}, will capture smeared phase transitions \cite{kuhn2017heterogeneous} but not necessarily their nature.

\subsection{Local susceptibility}

Finally, we propose a new measure of local susceptibility that contains some of the important information contained in the set of $P_i(p)$ curves. We want this local susceptibility to be a single curve describing the response function of how the spatial distribution of the local order parameter changes following a small variation in occupation probability. Since any uniform and linear combination of the $P_i(p)$ curves or their derivatives can be written as a function of the global order parameter, $S = \sum P_i(p) / N$, we know that averaging is unlikely to capture small localized responses. We instead look at a second order property of the $P_i(p)$ curve --- their standard deviation $\sigma \left(P_i(p)\right)$ --- which captures the heterogeneity of the $P_i(p)$ curves.

\begin{figure}[t!]
\centering
\includegraphics[width=0.9\linewidth]{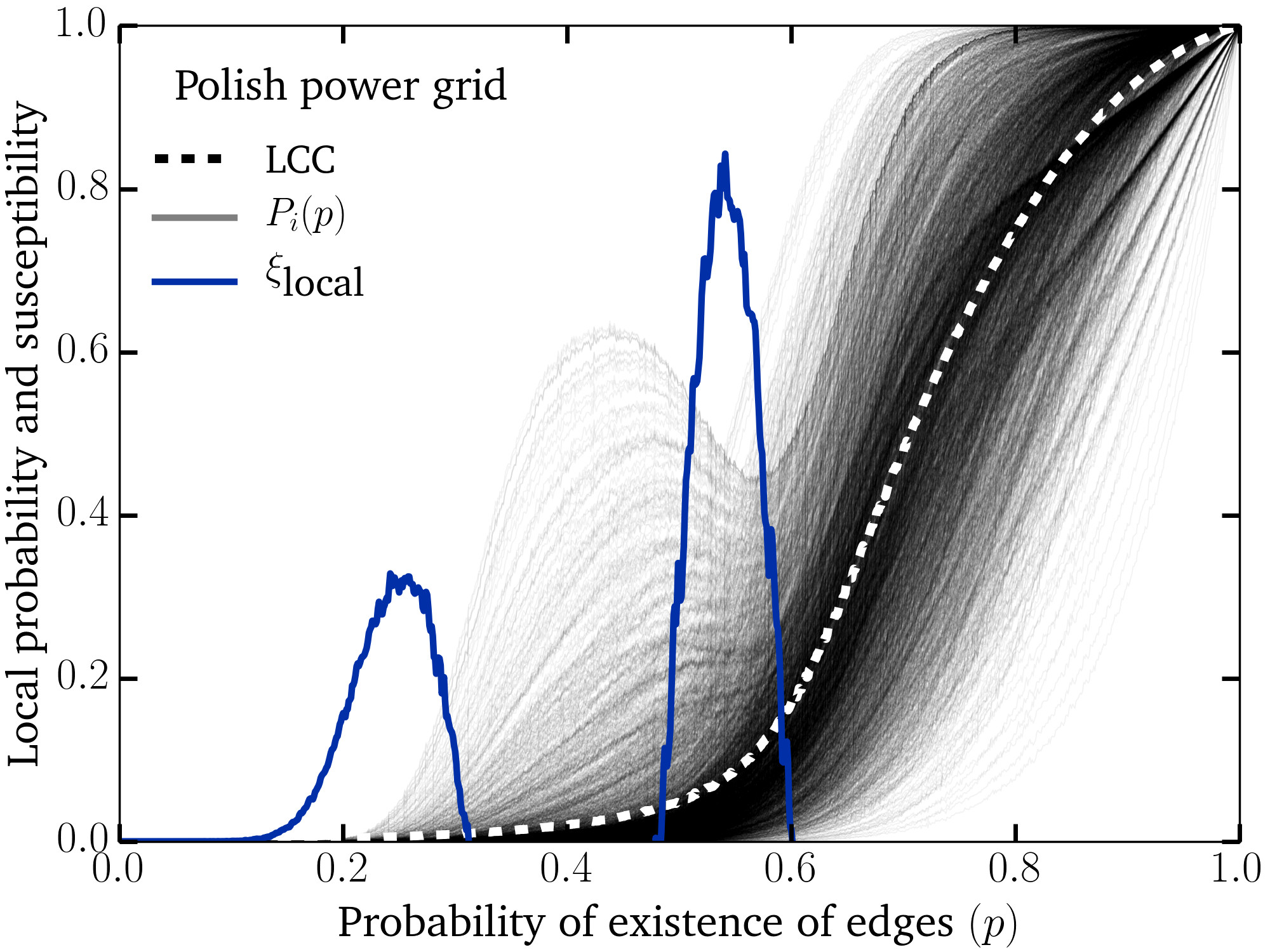}
\caption{\textbf{Using local susceptibility to detect smeared or sequential phase transitions.} The results of Eq.~(\ref{eq:localsusc}) in blue are compared to both the global and local order parameters (i.e. $S_1/N$ in white and $P_i(p)$ for all nodes $i$ in grey). Not only is the local susceptibility able to capture the first transition, unlike traditional susceptibility shown in Fig.~\ref{fig:detection}, but it also better detects the second transition around the point of maximum curvature in the global order parameter.  }
\label{fig:localsus}
\end{figure}

Local maxima of $\sigma \left(P_i(p)\right)$ correspond to points where some substructures have super-critical behavior while other may not, but that would not capture the initial percolation transition where all $P_i(p)$ are still close to zero. Local maxima of the first derivative in $p$ of $\sigma \left(P_i(p)\right)$ correspond to points where the spatial heterogeneity of the order parameter changes the most with varying $p$. And local maxima of the second derivative in $p$ of $\sigma \left(P_i(p)\right)$ correspond to points of maximum curvature where the response of the order parameter changes rapidly with varying $p$. These can be caused by both regular phase transitions, or sequential transitions in different subsystems. We thus propose

\begin{equation}
  \chi _{\textrm{local}} = \frac{d^2}{dp^2}\sigma \left(P_i(p)\right) \ ,
\label{eq:localsusc}
\end{equation}
where the derivative is to be performed numerically. Figure~\ref{fig:localsus} shows the local susceptibility and compares it with the global and local order parameters of the Polish power grid. As expected, the maxima of local susceptibility accurately capture the two main transitions.

\subsection{Conclusion}

In conclusion, we have observed that percolation transition in real complex networks can be often described through the lens of smeared phase transitions. We have illustrated with a few case studies that the nature of the inhomogeneities can be studied by looking at the distribution of local order parameters and determining whether subdivision of nodes by degree, modules, or centrality classes best explain the observed variations. We observed that modular and core-periphery structure can both be responsible for the observed smeared phase transitions and we discussed the qualitative difference between the two types of mesoscopic structure. Importantly, looking at the spatial distribution of the order parameter through the set of $P_i(p)$ curves can help identify the nature (or cause) of the smeared phase transition. And while in theory these results might all be due to the finite size of real networks, they are all captured by the state-of-the-art analytical approaches to percolation on complex networks. It is therefore an important feature to consider when comparing how well different analytical models predict the percolation transition. Similarly, it is critical to study the potential for smeared transition in real systems before applying results from clean phase transition to percolation-like models of system resilience or disease spread. In that regards, the proposed measure of local susceptibility can help identifying smeared or sequential phase transitions at a glance.\\

\begin{acknowledgments}
LHD acknowledges support from the National Science Foundations Grant No. DMS-1829826. AA acknowledges financial support from the the project Sentinelle Nord of the Canada First Research Excellence Fund, from ``la Caixa'' Foundation and from the Spanish ``Juan de la Cierva-incorporaci\'on'' program (IJCI-2016-30193). The authors also thank Guillaume St-Onge, Tiago Peixoto, Jean-Gabriel Young and Ginestra Bianconi for useful discussions and suggesting relevant literature.
\end{acknowledgments}

%
%
%
%
%
\end{document}